\documentstyle[12pt,moriond,psfig]{article}
\newcommand{\ergs}{erg~s$^{-1}$}

\newcommand{\ergcmds}{erg~cm$^{-2}$~s$^{-1}$}

\newcommand{\Msun}{\ifmmode M_{\odot} \else M$_{\odot}$\fi}
\newcommand{\Rsun}{\ifmmode R_{\odot} \else R$_{\odot}$\fi}
\newcommand{\Lsun}{\ifmmode L_{\odot} \else L$_{\odot}$\fi}
\newcommand{\Zsun}{\ifmmode Z_{\odot} \else Z$_{\odot}$\fi}

\newcommand{\qh}{\ifmmode Q_{H^{o}} \else $Q_{H^{o}}$\fi}
\newcommand{\qhe}{\ifmmode Q_{He^{o}} \else $Q_{He^{o}}$\fi}

\newcommand{\Tstar}{\ifmmode T^{\star} \else $T^{\star}$\fi}
\newcommand{\Lstar}{\ifmmode L^{\star} \else $L^{\star}$\fi}
\newcommand{\Mstar}{\ifmmode M^{\star} \else $M^{\star}$\fi}
\newcommand{\Teff}{\ifmmode T_{eff} \else $T_{eff}$\fi}
\newcommand{\Teq}{\ifmmode T_{eq} \else $T_{eq}$\fi}
\newcommand{\Mup}{\ifmmode M_{up} \else $M_{up}$\fi}
\newcommand{\Mlow}{\ifmmode M_{low} \else $M_{low}$\fi}
\newcommand{\Mneb}{\ifmmode M_{neb} \else $M_{neb}$\fi}
\newcommand{\Mion}{\ifmmode M_{ion} \else $M_{ion}$\fi}
\newcommand{\vexp}{\ifmmode v_{exp} \else $v_{exp}$\fi}

\newcommand{\hii}{H~{\sc ii} \ }

\newcommand{\Ha}{\ifmmode {\rm H}\alpha \else H$\alpha$ \fi}
\newcommand{\Hb}{\ifmmode {\rm H}\beta \else H$\beta$ \fi}
\newcommand{\Heii}{\ifmmode {\rm He~{\sc ii}} \lambda 4686 \else He~{\sc ii} $\lambda$4686 \fi}
\newcommand{\Nii}{\ifmmode {\rm [N{\sc ii}]} \lambda 6584 \else [N{\sc ii}] $\lambda$6584 \fi}
\newcommand{\nii}{\ifmmode {\rm [N{\sc ii}]} \else [N{\sc ii}]\fi}
\newcommand{\Oi}{\ifmmode {\rm [O{\sc i}]} \lambda 6300 \else [O{\sc i}] $\lambda$6300 \fi}
\newcommand{\oi}{\ifmmode {\rm [O{\sc i}]} \else [O{\sc i}]\fi}
\newcommand{\Oii}{\ifmmode {\rm [O{\sc ii}]} \lambda 3727 \else [O{\sc ii}] $\lambda$3727 \fi}
\newcommand{\oii}{\ifmmode {\rm [O{\sc ii}]} \else [O{\sc ii}]\fi}
\newcommand{\Oiii}{\ifmmode {\rm [O{\sc iii}]} \lambda 5007 \else [O{\sc iii}] $\lambda$5007 \fi}
\newcommand{\oiii}{\ifmmode {\rm [O{\sc iii}]} \else [O{\sc iii}]\fi}
\newcommand{\Oiiita}{\ifmmode {\rm [O{\sc iii}]} \lambda 4363 \else [O{\sc iii}] $\lambda$4363 \fi}
\newcommand{\Ciii}{\ifmmode {\rm C{\sc iii}]} \lambda 1909 \else C{\sc iii}] $\lambda$1909 \fi}
\newcommand{\ciii}{\ifmmode {\rm C{\sc iii}]} \else C{\sc iii}]\fi}
\newcommand{\Sii}{\ifmmode {\rm [S{\sc ii}]} \lambda 6725 \else [S{\sc ii}] $\lambda$6725 \fi}
\newcommand{\sii}{\ifmmode {\rm [S{\sc ii}]} \else [S{\sc ii}]\fi}
\newcommand{\Siii}{\ifmmode {\rm [S{\sc iii}]} \lambda 9532 \else [S{\sc iii}] $\lambda$9532 \fi}
\newcommand{\siii}{\ifmmode {\rm [S{\sc iii}]} \else [S{\sc iii}]\fi}
\newcommand{\Siiita}{\ifmmode {\rm [S{\sc iii}]} \lambda 6312 \else [S{\sc iii}] $\lambda$6312 \fi}
\newcommand{\Aiii}{\ifmmode {\rm [A{\sc iii}]} \lambda 7136 \else [A{\sc iii}] $\lambda$7136 \fi}
\newcommand{\rOiii}{\ifmmode {\rm [O{\sc iii}]} \lambda 4363/5007 \else [O{\sc iii}] $\lambda$4363/5007 \fi}

\newcommand{\Hp}{\ifmmode {\rm H}^{+} \else H$^{+}$ \fi}
\newcommand{\Hep}{\ifmmode {\rm He}^{+} \else He$^{+}$ \fi}
\newcommand{\Hepp}{\ifmmode {\rm He}^{++} \else He$^{++}$ \fi}
\newcommand{\Op}{\ifmmode {\rm O}^{+} \else O$^{+}$ \fi}
\newcommand{\Opp}{\ifmmode {\rm O}^{++} \else O$^{++}$ \fi}
\newcommand{\Oppp}{\ifmmode {\rm O}^{+++} \else O$^{+++}$ \fi}
\newcommand{\Np}{\ifmmode {\rm N}^{+} \else N$^{+}$ \fi}
\newcommand{\Npp}{\ifmmode {\rm N}^{++} \else N$^{++}$ \fi}
\newcommand{\Nppp}{\ifmmode {\rm N}^{+++} \else N$^{+++}$ \fi}
\newcommand{\Nepp}{\ifmmode {\rm Ne}^{++} \else Ne$^{++}$ \fi}
\begin{document}
\heading{THE ANALYSIS OF GIANT \hii REGIONS 
USING PHOTOIONIZATION MODELS}
 
\author{G. Stasi\'nska  DAEC, Observatoire de Meudon, France
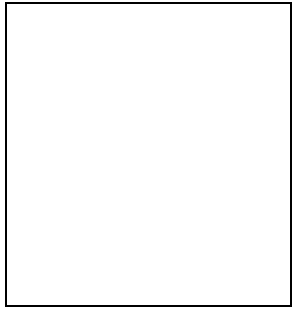
\footnote{
to appear in ``Dwarf Galaxies and Cosmology'', eds. Cayatte et al.,
 Editions Frontieres}}

\begin{moriondabstract}

After briefly reviewing the impact of various parameters on photoionization 
models, we illustrate the importance of the nebular geometry by two 
examples: we first compare the results from models with thin shell 
and full sphere geometry, and we study the influence 
of a diffuse ionized medium in the integrated spectra of galaxies at 
large redshift. Finally, we show how the calibration 
of the strong line method for deriving oxygen abundances depends on 
the assumed average properties of the ionizing star cluster and on the density 
distribution of the nebular gas. 

\end{moriondabstract}

\section{Introduction }
In the past, the analysis of giant \hii regions relied on single star 
photoionization models. Recent advances in stellar evolution and  population
 synthesis made it possible to build models of \hii 
regions that are ionized by star clusters (\cite {McG91},  \cite {McG94}, \cite 
{CMH94}, \cite 
{GV94}, \cite {GV95}, \cite {GSDal95}, \cite {SL96}).
 Such models are helpful in the quest 
of the best diagnostics of giant \hii regions and of the populations of 
hot stars embedded in them \cite {S96}. 

After a short introduction to photoionization modelling, we will mainly 
address two points. One concerns the influence of the nebular density 
structure. The other concerns the strong line method 
for deriving elemental abundances. This method has already been widely 
discussed in the literature, but deserves further attention because 
it is the only possible way to 
derive abundances in low surface brightness systems as well as in 
distant galaxies.

\section{Photoionization modelling in brief}

The primary parameters defining a model are (i) the oxygen abundance 
(oxygen usually provides 
most of the cooling and can be 
considered as the 
element defining the metallicity Z), (ii) the mean effective 
temperature $<T_{eff}>$ of 
the
ionizing radiation field (to first order, it can be characterized 
by the ratio \qhe / \qh \,  of the total number of photons above 24.6 
and 13.6 eV, and (iii) the ionization parameter $U_{S}$ whose definition 
for a spherical nebula of constant density is $U_{S} = \qh / 
(4\pi R_{S}^{2}n c)$ where $R_{S}$ is the Str\"omgren radius,  $n$ is the gas 
density and $c$ the speed of light. 

The intensities of the emission lines are proportional to the 
ionic populations of the elements producing them, but are strongly 
modulated by the electron temperature. The electron temperature 
results from a balance between the energy gains (due to ionization of 
hydrogen and helium, and essentially fixed by the 
effective temperature) and the energy losses (mainly due to 
line emission from the "metals", i.e. the elements heavier than H 
and He, except for the lowest metallicities where hydrogen losses become 
dominant).
The ionization parameter sets the 
distribution of the various ions within the nebula, once the radiation field
 is specified.

Among the secondary parameters are: (i) the relative abundances of the elements 
in the gas 
with respect to oxygen, (ii) the possible presence of dust 
and its composition, (iii) the details of the
energy distribution of the stellar ionizing photons, and (iv) the 
gas density distribution and the location of the ionizing stars with 
respect to the gas. 

The species whose abundance variations affect the 
thermal balance most are nitrogen, carbon, and the refractory elements. 
The N/O and C/O ratios are expected to vary from one object 
to another since nitrogen and carbon are not produced in the same nucleosynthetic 
sites as oxygen. If abundant, nitrogen and carbon contribute significantly to 
the cooling of the gas.  Refractory elements like Si, Mg, Fe may be depleted 
in the gas phase if 
dust is present, and deprive it from a cooling 
source which becomes important at high metallicities \cite {H93}. 

The direct effects of dust (apart from reddening) are 
mainly to compete with gas for the absorption of photons \cite {M86} but,  as shown 
by \cite {Bal91}, dust 
also participates in the heating or the cooling of the gas.
These effects strongly depend on the 
physical conditions inside the nebula and are difficult to model 
in detail, but they are of secondary importance in the thermal balance.

The energy distribution of the ionizing radiation field is dictated 
by the population of 
ionizing stars and the properties of their atmospheres. Even for a 
{\it given} mean
 effective temperature (or a 
mean \qhe / \qh), the stellar energy 
distribution influences the emission line properties 
 by affecting the ionization structure of the various 
elements \cite {SS97}. A review of 
the impact of stellar evolution and atmospheres can be found in \cite 
{GV96}. 

The effect of the distribution of the stars with respect to the gas 
has been considered in a simplified way by a few authors (\cite {Pe86}, 
\cite {BP97}) who modelled giant \hii regions as aggregates of 
single star Str\"omgen spheres rather than entities ionized 
by a dense star cluster located in their center, as is done in most 
studies. There is an increasing number of objects (e.g. 30 Doradus \cite {Hal95},
NGC 1569 \cite {DMal97},  
I Zw 18 \cite {HT95}) where 
the position of individual stars is known, and different situations exist.

So far, the influence of the density distribution has not been much
 studied. This partly stems from the fact that most photoionization 
codes assume spherical symmetry \cite {Fal95}. However, even in such a case, there 
is room for exploration, as seen below.

\section{The effect of the density structure on emission line ratios}

Here, we will briefly explore several 
simple geometries, using models obtained 
with the photoionization code PHOTO and a synthetic evolving starburst as 
described in  Stasi\'nska \& Leitherer \cite {SL96}. A more extended grid 
of models than the one 
presented in that paper has now been 
constructed. The details of all the models, including the intensities of 
about a hundred of lines from the far IR to the UV and other helpful 
parameters are available by anonymous ftp from 
ftp.obspm.fr/pub/obs/grazyna/cd-crete.

\subsection {Filled sphere versus hollow shell}

Two geometries have been mostly used to model giant \hii regions. One 
is a full sphere of constant density (e.g. \cite {SL96}) or uniformly 
filled with gas clumps of constant density  (e.g. 
\cite {McG91}). It has been shown on various occasions that any combination 
of \qh, $n$, and the filling factor $\epsilon$ keeping
 the ionization parameter $U_{S}$ constant results in the same nebular ionization 
structure.  A few authors have used another geometry, 
namely a hollow shell (e.g. \cite 
{GV95}). This choice is motivated by the ring structures  often 
seen in giant \hii regions in nearby galaxies (\cite {CL90}, \cite {OM94},
 \cite {Yal96}) 
and by the idea that 
the interaction of stellar winds and supernovae 
explosions from the central star cluster with the ambient interstellar gas creates large bubbles 
\cite {CMcCW75}, \cite {McCK87}. 

 \begin{figure}
\centerline{\psfig{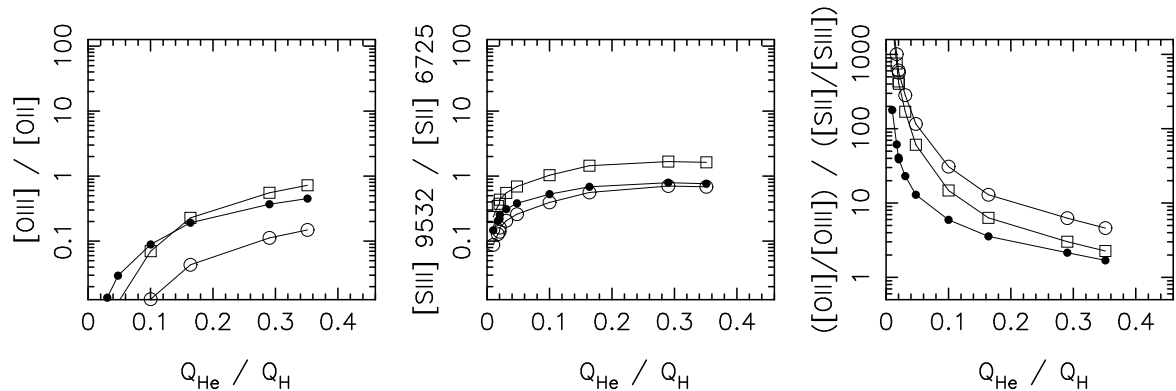}}
\caption{{\protect\small 
Emission line ratios as a function of \qh / \qhe \ in sequences of 
photoionization models for evolving starbursts at Z = 0.1 \Zsun \ that are identical in all 
respects except the geometry. Filled circles: full sphere 
models of uniform gas density; empty circles: hollow sphere models of same  $U_{S}$ as the 
full sphere models; empty squares: hollow sphere models of same $\bar{U}$ as the 
full sphere models. 
}}
\label{fig_HM_GEOM-.PS}
\end{figure}

In order to illustrate the role played by 
geometry, we compare in Fig. 1  several line intensity ratios as a 
function of \qhe/\qh \, for three sequences of evolving models of giant \hii regions
identical in every respect except the geometry. 
 Filled circles represent full sphere 
models. Empty circles represent hollow sphere models of same  $U_{S}$ . 
Empty squares represent hollow sphere models of same volume averaged 
ionization parameter $\bar{U}$
 as the full sphere models (as mentioned by \cite 
{McG91}, for a full sphere of constant density and given filling 
factor, one has $\bar{U}=3U_{S}$). It is clearly seen that the 
emission line spectra are different in all three models. At a given 
\qhe/\qh \ ,
the \oiii / \oii \ or \siii / \sii \ ratios are not simply dependent of the "ionization 
parameter" (whatever definition one takes), and in the example shown 
in Fig. 1, the dependence on the geometry is not the same for the two 
line ratios. As a consequence, the "radiation 
softness parameter"
(\oii/\oiii) / (\sii/\siii) used for ranking the 
mean effective temperature of \hii regions and believed to be 
relatively insensitive to ionization conditions  \cite {VP88}, does depend on 
the geometry.

 Unfortunately, there is
 no easy rule to predict the magnitude and even the direction of the changes 
when altering the geometry, even in such simple cases as considered 
here. Therefore, unless something is known about the geometry of the 
\hii regions one is studying, it is highly recommended to test the 
robustness of the inferences derived from photoionization mmodels 
by considering several exemplary geometries.

It is also important to realize that, in ab initio theoretical models of 
evolving starbursts, the prescription 
adopted for the geometry has an incidence on the evolution of the 
emission line intensity ratios.

Garc\'{\i}a-Vargas et al. \cite {GV95} adopted a 
thin shell geometry defined by a distance to the central star 
cluster held constant during the whole evolution of the cluster. 
In such a circumstance, we have $\bar{U}\propto \qh(t)$. Stasi\'nska 
\& Leitherer \cite {SL96} adopted a full sphere geometry of uniform density, 
and in such a case it is easily shown that
 $\bar{U}\propto (\qh(t))^{\frac{1}{3}}$.
That is, the ionization parameter decreases with \qh (and therefore 
with time) at a much 
slower rate than in the models of Garc\'{\i}a-Vargas et al. Clearly, 
any inference on starburst ages based on such models strongly depends 
on the adopted nebular properties. So, one cannot simply use the 
ionization parameter as a measure of the age of a given starburst, as 
suggested by Garc\'{\i}a-Vargas et al. Actually, a starburst can 
possibly be unveiled using emission line intensities of the 
surrounding \hii region only if additional
 information is available.

Note that, if, adopting a thin shell model, the radius were taken to 
vary with time like in the theory of stellar wind bubbles (\cite{D79}), one would 
have
$\bar{U}\propto t^{-\frac{6}{5}}\qh(t)$.  On the other hand, adopting a full 
sphere geometry, if gas were supposed to expand at constant velocity, 
one would have $\bar{U}\propto t^{-1}(\qh(t))^{\frac{1}{3}}$.  Of 
course, probably none of these two descriptions is realistic, although 
it is likely that the the ionization parameter decreases with time not 
only on the account of a decrease in \qh \, but also because of ram 
pressure effects and gas dissipation.

\subsection{Integrated spectra of core-halo \hii regions}

There is ample documentation that many giant \hii regions exhibit a 
core-halo structure. There is also evidence, in galaxies,
 for the existence of a diffuse 
ionized medium (\cite {R96}, \cite {WHL97}, \cite {M97}, \cite {MK97}). This diffuse gas is probably
ionized by photons leaking out from the core \hii regions, although 
other mechanisms may be at work in addition \cite {M97}.
 In such a picture, the core is either density bounded, at least in some 
directions, or ionization bounded with a covering factor 
less than one. The diffuse medium has a much lower ionization 
parameter than the core, therefore it emits strongly in \sii \ and 
\oi \ and weakly in \oiii. When studying a nearby giant \hii region, 
the diffuse ionized medium will hardly contribute to the observed 
spectrum. On the other hand, when observing a galaxy at high 
redshift, the slit will encompass the bright core together with the diffuse 
halo, and 
the resulting spectrum will be very different from that of an \hii 
region in a nearby galaxy \cite {LH94}. This "aperture effect" can be mistaken for 
a sign of "activity" or shock excitation in galaxies at large redshifts. 
An example of such a situation is shown in 
Table 1 and in Fig. 2.

\begin{figure}
\centerline{\psfig{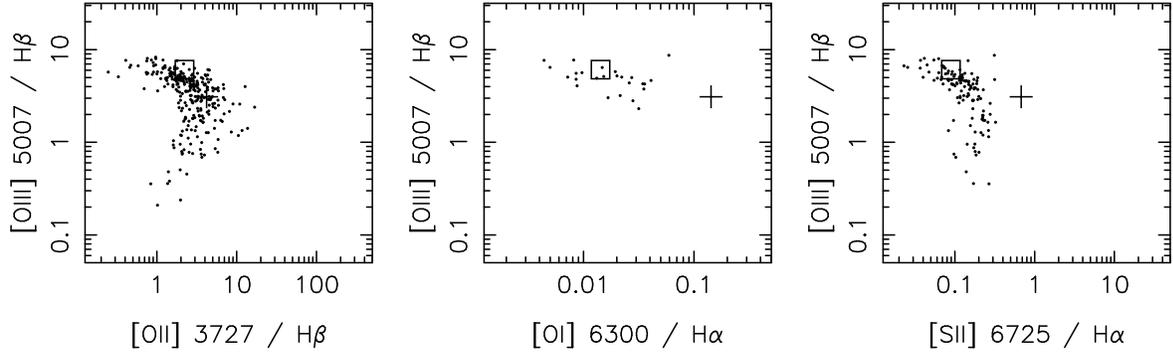}}
\caption{{\protect\small 
Diagnostic diagrams commonly used to detect signs of activity in 
giant \hii regions and in emission line galaxies. The dots represent the 
giant \hii  
regions observed in nearby galaxies by Mc Call et al. \cite {MCRS85}. 
The square and the cross show the position the model giant \hii regions
described in Table 1. The square corresponds to the 
bright core,
the cross corresponds to the integrated spectrum of a "core + halo"
 structure. In the \oiii / \Hb versus \oi / \Ha and \oiii / \Hb versus 
 \sii / \Ha diagrams, the core + halo is shifted 
 outside the \hii region sequence
 towards the domain of liners.
}}
\label{fig_heii}
\end{figure}

\setcounter{table}{0}

\begin{table*}
\caption{observing a core-halo \hii galaxy at different redshifts}
\begin{center}
\begin{tabular}{lll}
\hline
 \multicolumn{3}{c} {starburst parameters: 10$^{6}$\Msun, Z = \Zsun /4, 
  (\qh = 7.7 10$^{52}$~s$^{-1}$)} \\
\hline           

 \, & {\em core} & {\em halo} \\

n & 100 (cm$^{-3}$)& 1  (cm$^{-3}$)\\
$\epsilon$ & 10$^{-2}$ & 10$^{-4}$ \\
$U_{S}$ & 2.5 10$^{-3}$ &  2.5 10$^{-5}$ \\
$R_{S}$ & 300 pc & 30 Kpc  \\
\Mion (50\% covering) & 2 10$^{6}$~\Msun & 2 10$^{8}$~\Msun \\
\L(\Hb) (50\% covering) & 2 10$^{40}$~\ergs &  2 10$^{40}$~\ergs \\
\hline
 \multicolumn{3}{c} {at 10 Mpc (z=0.002)} \\
\hline
angular radius $\theta$  & 6" & 600" \\ 
F(\Hb) &  2 10$^{-12}$ (\ergcmds) & 2 10$^{-12}$ (\ergcmds) \\
\Oii/\Hb & 2.22 & 6.24 \\
\Oiii/\Hb & 6.09 & 0.16 \\
\Oi/\Hb & 0.04 & 0.77 \\
\Sii/\Hb & 0.25 & 3.48 \\
\hline
  \multicolumn{1}{c} {} & \multicolumn{2}{c} {at 1000 Mpc (z=0.2)} \\
\hline
angular radius $\theta$  & 0.06 " & 6 " \\ 
F(\Hb) &  2 10$^{-16}$ (\ergcmds) & 2 10$^{-16}$ (\ergcmds) \\
\, & \multicolumn{2}{l} {\ \ \ \ \ \ \ \ \ \ \ \ \ {\em core + halo}} \\
\Oii/\Hb & \multicolumn{2}{l} {\ \ \ \ \ \ \ \ \ \ \ \ \ \ \ \ \ \ \ 4.2} \\
\Oiii/\Hb & \multicolumn{2}{l} {\ \ \ \ \ \ \ \ \ \ \ \ \ \ \ \ \ \ \ 3.1} \\
\Oi/\Hb & \multicolumn{2}{l} {\ \ \ \ \ \ \ \ \ \ \ \ \ \ \ \ \ \ \ 0.4} \\
\Sii/\Hb & \multicolumn{2}{l} {\ \ \ \ \ \ \ \ \ \ \ \ \ \ \ \ \ \ \ 1.9} \\
\hline
\end{tabular}
\end{center}
\end{table*}

\section{Abundances from the strong line method}

\begin{figure}
\centerline{\psfig{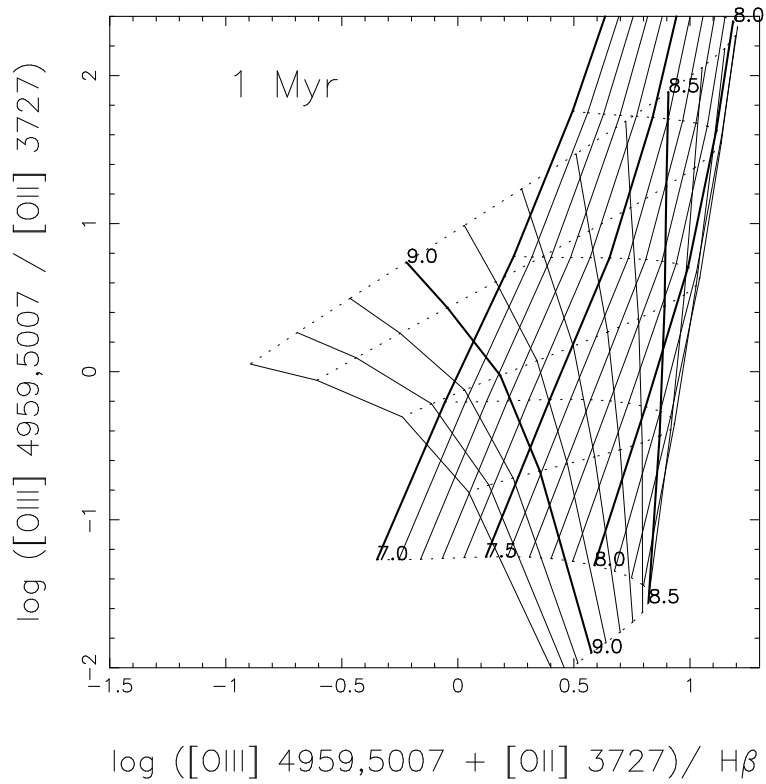}}
\caption{{\protect\small 
Mc Gaugh's diagram for a giant \hii region ionized by a
 1~Myr old star cluster as computed using the full sphere photoionization 
models of \cite {SL96}. The continuous lines correspond to loci of 
same O/H by intervals of 0.1 dex in log~(O/H) and are labelled with 
the value of log~(O/H)+12. The dotted lines correspond to loci of same 
total mass of the starburst, by intervals of 3~dex (corresponding to 
1~dex in $U_{S}$).
}}
\label{GR_FS1--.PS}
\end{figure}

\begin{figure}
\centerline{\psfig{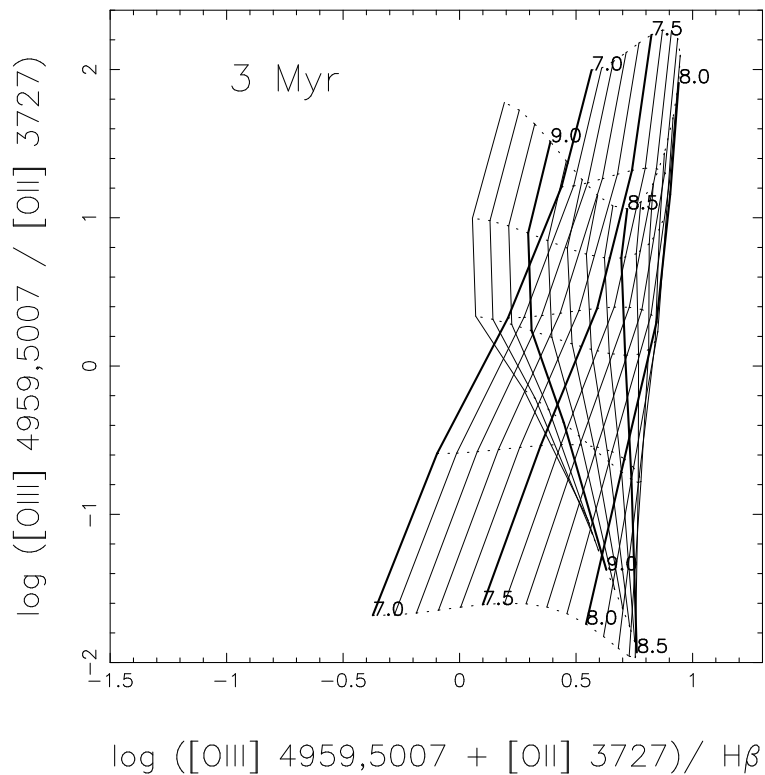}}
\caption{{\protect\small 
Same as Fig. 3, but for a 3~Myr old star cluster.
}}
\label{GR_FS3--.PS}
\end{figure}

\begin{figure}
\centerline{\psfig{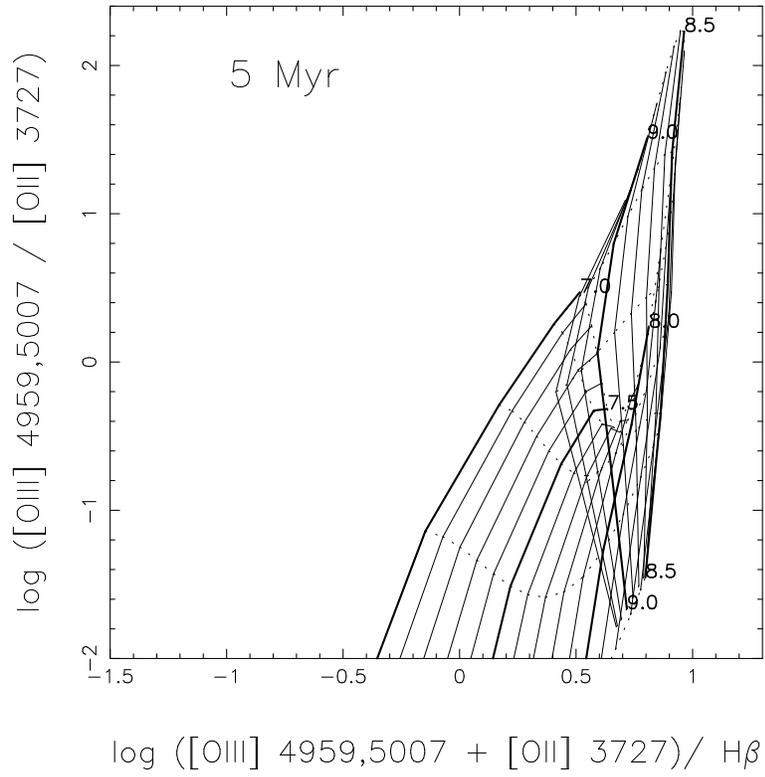}}
\caption{{\protect\small 
Same as Fig. 3, but for a 5~Myr old star cluster.
}}
\label{GR_FS5--.PS}
\end{figure}

\begin{figure}
\centerline{\psfig{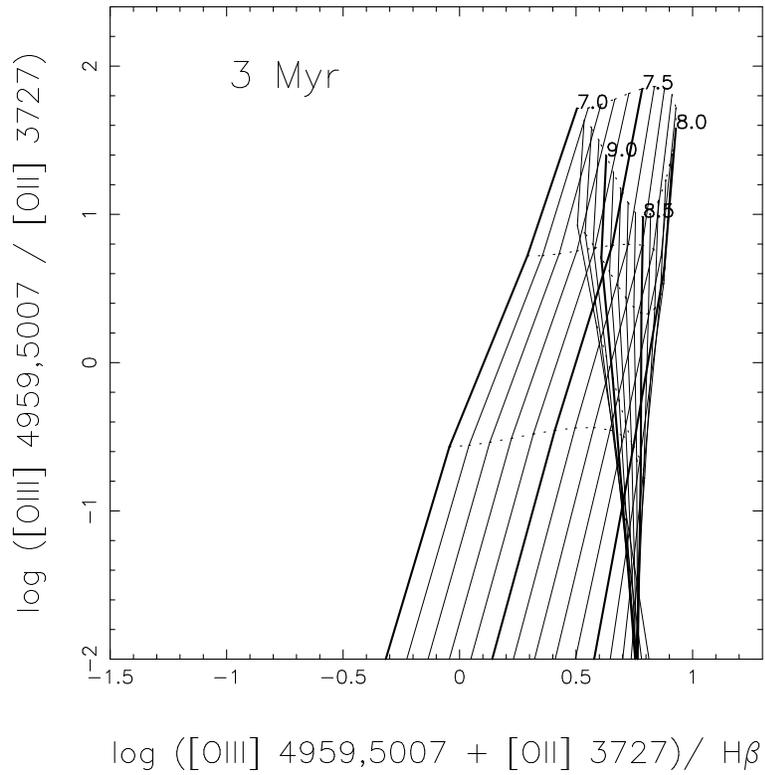}}
\caption{{\protect\small 
Same as Fig. 4, but for a different nebular geometry: hollow spheres.
}}
\label{GR_HS3-.PS}
\end{figure}

While abundances in \hii regions are readily obtained from line 
intensity ratios using 
electron temperature based empirical methods (the problem of 
temperature fluctuations should be kept in 
mind however, see  \cite{P95}, \cite {M95}, \cite {S98} and references therein), 
these methods cannot be used for distant or low surface brightness galaxies.
 Indeed, they require 
a measurement of the faint \Oiiita line. Even in close by \hii 
regions,
they cannot be used if the metallicity is higher than about half 
solar, because the electron temperature is then too low for \Oiiita
to be emitted. An interesting alternative to overcome this problem has been 
proposed by Pagel et al. \cite {Pal79}, using only the strongest 
lines. Initially, the method was intended to allow the determination 
of abundances in the metal rich giant \hii 
regions in the inner parts of spiral galaxies. This method has been further 
discussed in a number of papers (e.g. \cite  {MCRS85}, \cite {DE86}). 
Then, 
Skillman \cite {Sk89} showed its potential use for low abundance 
systems. Mc Gaugh (\cite {McG91}, \cite {McG94}) provided the first 
two-dimensional calibration taking explicitely into 
account the role of the ionization parameter. 

The leading idea is that, with two line ratios (\oii / \Hb  and 
\oiii / \Hb ) it is possible to estimate the three main defining 
parameters of an \hii region (O/H, $<T_{eff}>$  and $\bar{U}$) if there is 
a universal link between these parameters. 
And indeed, it seems that the initial mass function of 
the ionizing stars of giant \hii regions is universal (see Leitherer, 
this volume). This implies that, on average, one expects $<T_{eff}>$ and O/H 
to be related in such objects. 

Mc Gaugh has built a grid of photoionization models of various 
metallicities and ionization parameters ionized by clusters 
of OB stars, and provided a convenient diagram 
 to read out the values of $\bar{U}$  and O/H 
from \oiii \ / \oii \ versus 
(\oiii + \oii ) / \Hb. One problem is that the relation 
between (\oiii + \oii ) / \Hb and O/H is double valued: at 
low metallicities, the intensities of \oii \  and \oiii\  with 
respect to \Hb increase with increasing O/H. But when the oxygen 
abundance gets high enough for this element to become the main coolant, 
the electron temperature drops and the cooling is gradually 
transferred from the optical 
forbidden lines to the infra red fine structure lines. Consequently, 
(\oiii + \oii ) / \Hb becomes a decreasing function of O/H. 
If nothing is suspected a priori about the metallicity of an
object, one can use \nii \ / \oii \  as a 
discriminant between the two regimes: this ratio is expected to be small ($<$~0.1) at low 
metallicities \cite {McG94}.

Mc Gaugh analyzed various sources of errors in the method and estimated an 
intrinsic uncertainty of 0.05~dex in log (O/H) at the 
low abundance end, 0.1~dex at the high abundance end, and 0.2~dex in 
the turn over region. The expected overall
uncertainties in the oxygen abundance are obtained by combining these
formal error bars with those due to observational errors \cite 
{McG94}, giving an estimated accuracy of about 0.2~dex for reasonably 
high signal-to-noise data, and worse in the turn over region.

Note that, on the low abundance end, 
comparing the values of O/H derived from the strong line method and 
from the electron temperature based method provides a direct estimate of 
 the error. Unfortunately, this cannot be done on 
 the high abundance end. Even tailored 
 photoionization models of metal rich \hii regions fitting all the 
 important lines do not provide a reliable calibration: the observational 
 constraints are generally far from sufficient to allow a precise 
 determination of the oxygen abundance in  metal rich objects where,
 as developed below, the emission line intensities 
 are extremely sensitive to the physical conditions in the nebulae.

Mc Gaugh's computations were done assuming a zero age 
starburst of given upper stellar mass limit, \Mup. The evolution of 
a starburst was mimicked by considering different values of \Mup.
Mc Gaugh's diagram is constructed for  $\Mup = 60~\Msun$, corresponding to 
an average age of $\sim$ 3~Myr for a burst having initially  $\Mup = 100~\Msun$.

Using evolving synthetic starburst models \cite {LH95}, 
it is  possible to follow the distortion of Mc Gaugh's diagram as the 
starburst ages. This is done in Figs. 3, 4 and 5, which show the 
diagram obtained by bicubic spline interpolation on the grid of 
photoionization models of \cite {SL96} at 1~Myr, 3~Myr and 5~Myr 
respectively.  The continuous lines correspond to loci of same O/H, 
while the dotted lines correspond to loci of same total initial mass 
of the stars (and not same ionization parameter as in Mc Gaugh). 

We see that, on the low abundance end, the main effect of ageing 
 is a drop in \oiii / \oii, which is due to a decrease of
 $\bar{U}$ implied by the lowering of the number of ionizing photons. The isoabundance curves
are not much displaced, especially at low excitation. Indeed, in this 
regime, cooling is mainly due to collisional excitation of H
Ly $\alpha$ and is a steep function of the electron temperature. 
Therefore, the electron temperature is insensitive to changes 
in the heating rate, and so are the intensities of the \Oiii  and 
\Oii. As a consequence, the oxygen abundance derived from such
 a diagram is not too dependent on age. A typical value of the 
 corresponding uncertainty is $\pm$ 0.1~dex 
 for an age of  3 $\pm$ 1~Myr.

On the high abundance end, the distortion of Mc Gaugh's diagram 
 is impressive. At high metallicities, 
cooling is provided by the infrared fine structure lines 
[O{\sc iii}] $\lambda$$\lambda$52, 88$\mu$, which are insensitive to the electron temperature. 
Therefore, the electron temperature is extremely affected by changes 
in the energy gains, and so is the intensity of  \Oiii. 
To make things worse, at high metallicities, the mean effective 
temperature of the stellar radiation is subject to large variations 
during the evolution (see Table 3 of \cite {SL96}). An 
important 
hardening occurs when stars reach the Wolf-Rayet phase, 
boosting the \Oiii line. As a result, oxygen abundances 
derived from the strong line method are quite uncertain at the 
high metallicity end. Typically, an \hii region having log (O/H)+12 
= 9 and an age of 5 Myr would appear less metal rich by 0.3~dex  
if interpreted with a diagram corresponding to 3~Myr. Other 
sources of uncertainties contribute as well, owing to the fact that the electron temperature 
is so ill defined. For example, any density irregularity, any change 
in the geometry will greatly affect the emission in \Oiii. 
An example is shown in Fig. 6, which presents Mc Gaugh's diagram at 
 3~Myr like Fig. 4, but for hollow spheres of same $\bar{U}$ as 
the full sphere
models used in Fig. 4. Note that Figs. 3 -- 6 were constructed by 
extrapolating the solar metallicity models towards higher abundances. 
We have also constructed models at twice solar metallicity (they are 
available on the ftp site mentioned above). Using these models produces 
very twisted isoabundance curves even around log (O/H)+12 = 9, because at 
metallicities twice solar, the strong cooling may completely quench 
the \Oiii \ line. The diagram is 
further complicated by the fact that, at high metallicities,  
recombination takes over
collisional excitation in the production of the \oii \ line.

As regards the turnover region, warnings have already been expressed by 
several authors. 
 In this regime, the main cooling agent is \Oiii, 
therefore the intensity of this line is independent of 
O/H, and is a function of the heating rate only. Nevertheless, one 
can say that an \hii regions showing (\oiii + \oii ) / \Hb $>$ 0.7 
must have a log (O/H)+12 within a few dex of 8. 

In summary, the strong line method  is 
rather robust at low abundances (log (O/H)+12 $<$ 7.7) 
 but is rather uncertain at higher 
metallicities. The fact that in spiral galaxies, one does see a trend 
in  (\oiii + \oii ) / \Hb with galactocentric radius (e.g. \cite 
{KG96}) is probably indeed related to an abundance effect. In this 
case, observing a large number of \hii regions at a given 
galactocentric radius will smooth out the effects due to
non uniformity of the stellar populations. Still, the unknown 
geometry of the \hii regions affects any calibration of the method. 
Trustworthy abundance determinations in 
metal rich \hii regions have to await the observation of 
the infrared fine structure lines. 

It has been suggested \cite {SMcW93} that the line ratio \Aiii / 
\Siii \ could be an indicator of metallicity in metal rich object, 
being an indicator of electron temperature. We do not share this 
optimistic view
since, as commented above, the electron temperature is ill 
defined at high metallicities and strongly dependent on the 
properties of the ionizing clusters and on the density structure of 
the nebulae.

The determination of the N/O ratio has been specifically discussed by several 
authors \cite {VCE93}, \cite {TEH96}, with the help of single star
photoionization models. Using the grid of models for evolving 
starburst presented in Sect. 3 as an interpolating device, 
one can obtain the N/O ratios from the observed \nii / \oii \ with a 
typical uncertainty of $\pm$ 0.15~dex.

%\noindent
%{\small {\sl Acknowledgments}
%GS is grateful to the ``GdR galaxies'' for 
%financial support. }

%
\vfill

\vspace*{-0.5cm}
\begin{moriondbib}
{\small 
\bibitem {Bal91}  Baldwin J.A., Ferland G.J., Martin P.G. et al, 1991, \apj 
{374} {580}
\bibitem {BP97} Barbaro C., Poggianti B.M., 1997, \aa {324} {490}
\bibitem {CL90} Chu Y.H., Mc Low M-M., \apj {365} {510}
\bibitem {CMcCW75} Castor J., McCray R., Weaver R., 1975, \apj {200}{L107}
\bibitem {CMH94} Cervi\~no M., Mas-Hesse J.M., 1994, \aa {284} {749}
\bibitem {DE86} Dopita M.A., Evans I.N., 1986, \apj {307} {431}
\bibitem {D79} Dyson J.E., 1979, \aa  {73} {132}
\bibitem {DMal97} De Marchi G., Clampin M., Greggio L. et al., 1997, \apj {479} {L27}
\bibitem {Fal95} Ferland G.J. et al., 1995, in {\it Analysis of 
Emission Lines: a Meeting Honoring the 70th Birthdays of D.E. 
Osterbrock and M.J. Seaton}, eds. R.E. Williams \& M. Livio (Cambridge 
Univ. Press), 83
\bibitem {GV94} Garc\'{\i}a-Vargas, M. L., D\'{\i}az, 1994, \apjs  {91} {553}
\bibitem {GV95}  Garc\'{\i}a-Vargas, M. L.,  Bressan, A., D\'{\i}az, A.I., 1995, 
\aas 
{112} {13}
\bibitem {GV96} Garc\'{\i}a-Vargas, M. L., 1996, in {\it From Stars
to Galaxies}, ASP Conf. Series, Vol. 98, eds. C. Leitherer, U. Fritze-v. Alvensleben,
and J. Huchra, 244 
\bibitem {GSDal95} Garnett D.R., Skillman E.D., Dufour R.J. et al., 
          \apj {443} {64} {76}
\bibitem {H93} Henry, R.B.C., 1993, \mnras {261} {306}
\bibitem {HT95} Hunter D.A., Thronson H.A. Jr., 1995, \apj {452} {238}
\bibitem {Hal95} Hunter D.A., Shaya E.J., Scowen P., et al, 1995, \apj {444} {758}
\bibitem {KG96} Kennicutt R.C., Garnett D.R., 1996, \apj {456} {504}
\bibitem {LH95} Leitherer C., Heckman T.M., 1995, \apjs {96} {9}
\bibitem {LH94} Lehnert M.D., Heckman T.M., 1994, \apj {426} {L27}
\bibitem {MCRS85} Mc Call M.L., Rybski P.M., Shields G.A., 1985, \apjs {57} {1}  
\bibitem {McCK87} Mc Cray R., Kafatos M.C., 1987, \apj {317} {190}       
\bibitem {McG91} Mc Gaugh S.S., 1991, \apj {380} {140}
\bibitem {McG94} Mc Gaugh S.S., 1994, \apj {426} {135}
\bibitem {M97} Martin C.L., 1997, \apj {491} {561}
\bibitem {MK97} Martin C.L., Kennicutt R.L., 1997, \apj {483} {698}
\bibitem {M86} Mathis J.S., 1986, {\em PASP \/} {\bf 98}, {995}}
\bibitem {M95} Mathis J.S.., 1995, {\em Rev Mex AA (Serie de 
Conferencias) \/} {\bf 3}, {207}
\bibitem {OM94} Oey M.S., Massey P., 1994, \apj {425} {635}
\bibitem {Pal79} Pagel B.E.J., Edmunds M.G., Blackwell D.E. et al., 
1979, \mnras {189} {95}
\bibitem {P95} Peimbert M., 1995, in {\it The Analysis of Emission Lines}, 
eds. Williams. R.E., Livio M. , Cambridge University Press p. 165
\bibitem {Pe86} Pe\~na M., 1986, {\em PASP \/} {\bf 98}, {1061}
\bibitem {R96} Rand R.J., 1996, \apj {462} {712}
\bibitem {SK95} Shields J.C., Kennicutt R.C., 1995, \apj {454} {807}
\bibitem {Sk89} Skillman E.D., 1989, \apj {347} 883
\bibitem {SL96} Stasi\'nska G., Leitherer C., 1996, \apjs {107} {661}
\bibitem {S96} Stasi\'nska G., 1996, in {\it From Stars
to Galaxies}, ASP Conf. Series, Vol. 98, eds. C. Leitherer, U. Fritze-v. Alvensleben,
and J. Huchra, 232 
\bibitem {S98} Stasi\'nska G., 1998, in {\it Abundance profiles: 
diagnostic tools for galaxy history}, ASP Conf. Ser., eds. Friedli et al.  in press
\bibitem {SS97} Stasi\'nska G., Schaerer D., 1997, \aa {322} {615}
\bibitem {SMcW93} Stevenson C.C., Mac Call M.L., Welch D.L., 1993, \apj {408} {460}
\bibitem {TEH96} Thurston T.R., Edmunds M.G., Henry R.B.C., 1996, \mnras 
{283} {990}
\bibitem {VCE93} Vila-Costas M.B., Edmunds M.G., 1993, \mnras {265} {199}
\bibitem {VP88} Vilchez J.M., Pagel B.E.J., 1988, \mnras {231} {257}
\bibitem {WHL97}Wang J., Heckman T.M., Lehnert M.D., 1997, \apj {491} 
{114}
\bibitem {Yal96} Yang H., Chu Y.H., Skillman E.D., Terlevich R., 1996, 
{\em Astronom. J. \/} {\bf 112}, {146}


\end{moriondbib}
\end{document}